\documentclass[lettersize,journal]{IEEEtran}
\usepackage{amsmath,amsfonts}
\usepackage{algorithm}
\usepackage{array}
\usepackage{textcomp}
\usepackage{stfloats}
\usepackage{url}
\usepackage{verbatim}
\usepackage{graphicx}
\usepackage{cite}
\usepackage[subpreambles=true]{standalone}

\usepackage[breaklinks=true]{hyperref}
\hypersetup{colorlinks=true, pdfstartview=FitV, linkcolor=black, citecolor=black, urlcolor=blue}

\usepackage{algpseudocode}
\usepackage{geometry}

\usepackage[format=plain,justification=centering]{caption}
\usepackage{subcaption}

\usepackage{booktabs}
\usepackage{color,soul}
\usepackage{rotating}

\usepackage{multirow}
\usepackage[table, xcdraw]{xcolor}
\usepackage{svg}

\usepackage{soul}

\definecolor{myblue}{rgb}{0.8, 0.85, 1}  
\definecolor{mygreen}{rgb}{0.85, 1, 0.85}  

\colorlet{rulecolor}{myblue}

\newcommand{\colorboxrule}[3][rulecolor]{
  \textcolor{#1}{\rule{#2}{#3}}
}

\usepackage{adjustbox}
\usepackage{tabu}

\graphicspath{ {./figures/} }

\usepackage[below]{placeins}
\usepackage{comment}
\usepackage{color}
\usepackage{geometry}
\usepackage{algpseudocode}

\usepackage{algorithmicx}
\usepackage{svg}

\usepackage[normalem]{ulem}

\usepackage{tikz}
\usepackage{enumitem}   
\usepackage{amssymb}    

\usepackage{orcidlink}

\usepackage{balance}

\newcommand{\checkedbox}{%
  \tikz[baseline=0.30ex]{
    \draw[line width=0.5pt] (0,0) rectangle (0.32,0.32) 
    node[pos=0.5]{\small\checkmark};
    }%
}

\begin{document}

\title{Distributed Service Orchestration in Edge-Cloud Continuum for Digital Healthcare}
\author{Johirul Islam\textsuperscript{\orcidlink{0000-0002-7523-0666}}, 
Hafiz Faheem Shahid\textsuperscript{\orcidlink{0009-0005-3350-0276}},
Ijaz Ahmad\textsuperscript{\orcidlink{0000-0002-6152-8947}}, 
\\
Tanesh Kumar\textsuperscript{\orcidlink{0000-0002-5907-8414}},
Ayan Mondal\textsuperscript{\orcidlink{0000-0003-1548-0580}}, and
Erkki Harjula\textsuperscript{\orcidlink{0000-0001-5331-209X}}
\thanks{Manuscript received April 19, 2026; revised August 16, 2026.}
}


\markboth{IEEE TRANSACTIONS ON PARALLEL AND DISTRIBUTED SYSTEMS,~Vol.~01234, No.~56789, August~2026}%
{Islam \MakeLowercase{\textit{et al.}}: Docker Private Registry}

\IEEEpubid{0000--0000/00\$00.00~\copyright~2026 IEEE}

\maketitle

\begin{abstract}
Today's digital healthcare services rely on various applications and functions that must be continuously accessible. Cloud computing enables global access to these services through public networks, which often also introduce increased latency, higher bandwidth consumption, and additional security risks compared to local operation. Edge computing mitigates these challenges by deploying cloud services closer to the end users and data sources, thereby improving resilience to network disruptions and reducing latency, bandwidth usage, and exposure to security threats. However, service deployment typically relies on the availability of centralized registry servers. Consequently, if the network connection or the registry server itself 
becomes unavailable, service deployment at the target edge node may fail. To address this, we propose a three-tier registry architecture to enhance deployment reliability and service availability, considering DockerHub as a remote public registry, an MEC-based off-premises registry as a remote private registry, and a LAN-based on-premises registry as a local private registry. The performance and efficiency 
are analyzed through measurements related to the estimated deployment time, inflicted network and computational load, and energy consumption, while the required nanoservices are deployed from different tiers. 
The experimental results indicate that alongside the improved tolerance to network disruptions, the local private and remote private registries also outperform the remote public registry in the deployment performance. The findings highlight the effectiveness of proximity-aware service distribution in improving the resilience, performance and efficiency of service deployment.

\end{abstract}

\begin{IEEEkeywords}
IoT, Edge-Cloud Continuum, Microservice Orchestration, Healthcare Applications.
\end{IEEEkeywords}

\section{Introduction}

\noindent The ever-growing demand for real-time processing and analysis capabilities in healthcare applications necessitates robust and latency-sensitive computing infrastructures \cite{9964122}. Cloud-based computing architecture was for long the reference architecture for most IoT applications. However, the original centralized model suffers from high latency and vulnerability to network disruptions. Edge–cloud architecture was introduced as a solution to address these challenges by integrating edge computing resources, located close to end users or sensor devices, with cloud infrastructure. \cite{Nuguri2026188}.
To further improve the system reliability, privacy protection, and resource efficiency, the authors in \cite{harjula2019decentralized, 10843208, 11392773} proposed a three-tier edge-cloud architecture combined with a distributed nanoservice architecture, which together made it possible to deploy service parts to constrained-capacity local devices. While this three-tier architecture successfully reduces the reliance of the already deployed services to the constantly available and high-performance underlying network architecture, one critical challenge remains: in most of the current systems, the service deployment relies on a central service registry \cite{Zerouali2018, Osorio2018}. This model compromises the reliability of the service deployment during, e.g., network or registry server outages, potentially hindering the deployment of critical healthcare or other applications when those are critically needed.

\IEEEpubidadjcol

In this paper, we address this critical limitation by proposing a three-tier service registry model to ensure the system's ability to deploy critical services in all circumstances, including when the connection to public service registries is down. In this paper, we exemplify such critical service with a medical first-response use case, where we implement a digital care pathway for an emergency patient requiring first aid at a scene, followed by monitoring the patient's vitals from the site during ambulance transportation to the hospital. This care pathway service is implemented based on patient monitoring and care-related nanoservices that are deployed or undeployed based on the need during the care pathway.
These aspects include the hardware requirements for local computing nodes, bandwidth requirements for the local networks, energy consumption on the local devices, and overall performance of the deployment in different deployment scenarios. Based on the gained knowledge, we are able to outline the best practices for, e.g., selecting which local nodes could be utilized as local edge nodes or which nanoservices should be maintained in the limited-capacity local registry, prioritizing the most critical, most often-used, and least resource-consuming services.
The key contributions of this paper are:
\begin{itemize}
    \item The concept for robust service orchestration to enable the seamless deployment of services in a medical use case.
    \item Three-tier service registry architecture to ensure the deployment of critical services during remote service registry unavailability.
    \item Evaluating the performance and efficiency of the proposed concept and analyzing the impact of the results in the selected use case scenario and beyond.
\end{itemize}

The rest of this paper is organized as follows: Section \ref{sec:background-and-related-works} discusses the background and related works. Section \ref{sec:usecase} and Section \ref{sec:system-model} present the selected use case and its system model, respectively. Section \ref{sec:implementation} and section \ref{sec:evaluations} discusses the implementation setup and evaluation results of the system model. Discussion and the future are presented in Section \ref{sec:discussion-and-future-works}. Finally, Section \ref{sec:conclusion} concludes the paper.

\section{Background and related works}
\label{sec:background-and-related-works}

\subsection{Evolution of edge-cloud computing} 
\noindent Cloud computing is a widely used concept for delivering applications, software platforms and computational resources over the Internet, instead of deploying those on dedicated servers or personal devices \cite{marinescu2022cloud, 9615028, 9403939}. This approach offers scalability, flexibility, and cost-efficiency as services can be scaled up or down based on user demand without the need for  infrastructure investments and maintenance. In healthcare domain, medical equipment generate vast amounts of health-related data, ranging from patient vital signs to diagnostic imaging. Cloud computing provides a centralized platform for storing, processing, and analyzing these data, facilitating real-time monitoring and data-driven decision-making \cite{9645306, Prabu2024277}. Additionally, cloud computing in medical IoT enables the implementation of sophisticated analytics and machine learning algorithms.
Cloud computing is, however, highly dependent on a reliable internet connection, and it introduces increased latency, higher bandwidth consumption, and also additional security risks compared to local operation. 

Edge computing is a distributed computing paradigm that involves processing data closer to the users and data sources, rather than relying solely on centralized cloud servers \cite{9301260, 10614854}. It reduces latency and enhances real-time processing capabilities, while improving robustness against network connectivity problems and security threats \cite{9301260, 10129089}. Shahid et al. \cite{11392773} has conducted a comprehensive comparative analysis of five architectural frameworks for IoT service orchestration in the edge–cloud continuum of 6G networks, namely traditional IoT–cloud, edge computing, fog computing, Multi-access Edge Computing (MEC), and local edge computing.
The edge-cloud approach is particularly relevant in healthcare, where reliable, efficient and timely processing of data from medical devices is crucial. Edge computing alleviates the load on the network by processing data through the resources in the user's proximity and sending only relevant data to the cloud \cite{Suriyan2025323, Hussain2024388}. This reduces not only the network congestion but also the costs associated with data transmission and storage, and helps limit unnecessary propagation of sensitive patient data \cite{10614854, Ramanathan20241, Karami20256183}.

\begin{figure*}[hb]
  \centering
  \includegraphics[trim=0.95in 0.25in 0.95in 0.29in,clip,width=0.9\linewidth,height=0.6\linewidth]{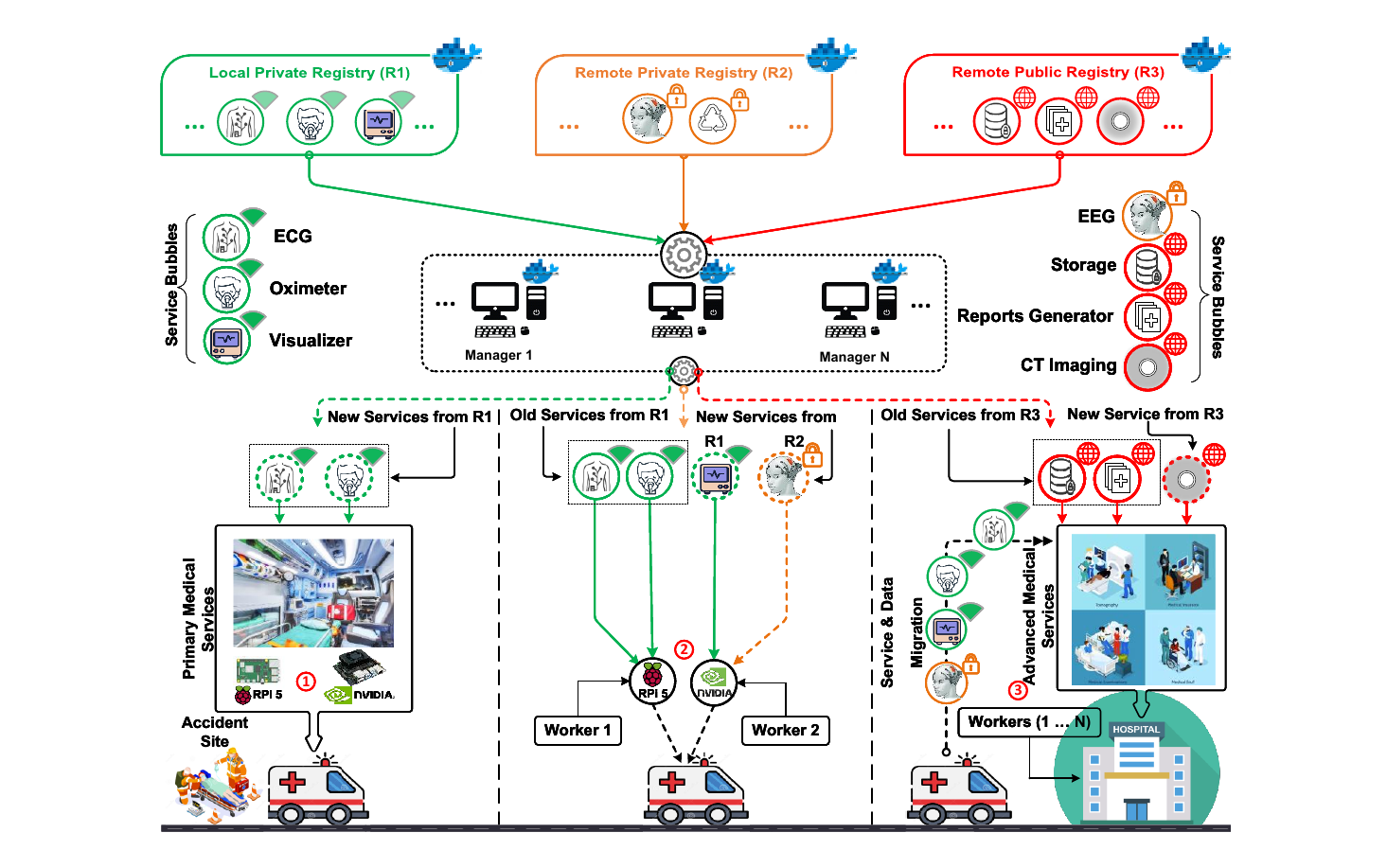}
  \caption{Resilient deployment of nanoservices for emergency patient monitoring.}
  \label{fig:resilient_deployment}
  \vspace{-10px}
\end{figure*} 

\subsection{Edge-cloud service orchestration}
\noindent Microservice architecture is a widely used decentralized approach that structures online applications as a collection of small, independent services. These services operate as autonomous entities and communicate with one another through well-defined APIs. Many of today's IoT computing nodes include sufficient processing power to process data and execute different actions locally. There is a clear need for mechanisms enabling harnessing this potential to practical operation. Harjula et al. have proposed a \textit{nanoservice} approach \cite{harjula2019decentralized, 8931321},  defining methods to deploy ultra-lightweight virtualized edge-cloud microservices on local computing devices, including IoT sensors and actuators. Docker swarm or Kubernetes are then used to deploy and orchestrate the containers across the cluster nodes.

Nanoservices extend the edge-cloud computing architecture to the local tier, enabling, e.g., local data pre-processing to save network bandwidth or avoid propagating sensitive data outside a room, floor, or building, thereby also improving privacy. Locally deployed functionalities also help addressing local network problems by enabling fully local operation of critical functions, thereby improving service resilience. To support this, Shahid et. al. \cite{shahid2024resource} have introduced a concept for intelligent resource-aware orchestration of nanoservices as a part of a comprehensive three-tier edge-cloud architecture. In \cite{11220483}, they proposed an AI-driven and ML-based heuristic algorithmic approach for the orchestration of IoT services by performing task scheduling, classification of each type of service, load balancing and optimizing the computing resource in the distributed edge-cloud continuum.

Although the concepts above enable the resilient operation of already deployed microservices and nanoservices, the current deployment methodologies remain heavily dependent on stable and continuously available network connectivity, as these services are typically provisioned from remote service registries.
This poses a major obstacle: if there are disruptions in the network, despite the reliable operation of already deployed services, new services cannot be deployed until the network operation has been restored. This dependence on network stability highlights a serious weakness in the current solutions, especially when it comes to healthcare, where service continuity is crucial \cite{wiig2021resilient}.
However, the next investigations should prioritize enhancing performance of constructing resilient systems capable of enduring network interruptions and guaranteeing ongoing healthcare service provision\cite{liu2021performance}. 

The emergence of distributed edge orchestration and service deployment has presented new opportunities for addressing the challenge of resilient service placement at the edge, such as presented in \cite{distributedorchestration} and \cite{9903191}. The former proposes a decentralized container scheduler capable of building an edge cloud from volunteer resources using Docker containers, while the latter introduces a dynamic, event-driven orchestration architecture leveraging decentralized registries and local decision logic. However, both works focus primarily on architectural design and proof-of-concept implementations.
In this paper, we aim to further advance the state-of-the-art by implementing a functional prototype for a real-life application scenario, and evaluate feasibility of the concept against the traditional centralized deployment, with respect to the above metrics.

\section{Use case: Emergency response scenario}
\label{sec:usecase}

\noindent
As the use case for our work, we consider an emergency response scenario, where a patient in a serious medical condition -- resulting from, e.g. a serious injury, stroke, etc. -- needs to be transported to a hospital. 
The emergency response team needs to start the primary emergency medical procedures already on site and extend the procedures during ambulance transportation to ensure the best patient outcome. 

Initially, they attach medical sensors to monitor the real-time health status of the patient, and during ambulance transportation, they enhance the monitoring with more advanced sensors and analytic tools that are connected to the hospital systems. Finally, the patient is handed over to the emergency unit of the hospital, during which the system must ensure the continuity of critical monitoring and treatment. The monitoring and treatment functions are implemented as nanoservices that follow the patient during the care pathway. Fig.~\ref{fig:resilient_deployment} presents the overview of the use case, whereas Fig.~\ref{fig:usecase_scenario} shows relevant phase-by-phase procedures that are used to deploy the use case nanoservices. Following are the four key phases identified for this emergency response use case:

\begin{figure*}[hb!]
  \centering
  \includegraphics[trim=0.25in 0.25in 0.25in 0.25in,clip,width=0.75\linewidth]{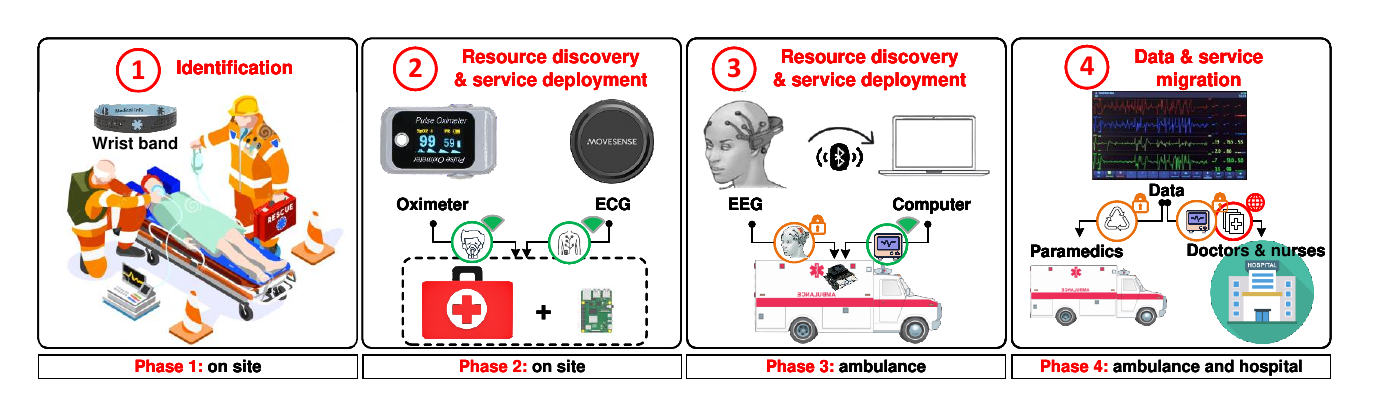}
  \caption{Phase-by-phase procedures that are used to deploy use case nanoservices.}
  \label{fig:usecase_scenario}
  \vspace{-10px}
\end{figure*} 

                
    
\textit{Phase 1 (on-site) -- Identification:} To monitor the patient with various medical sensors, the patient needs to be identified on site. For this, the paramedics attach a BLE-based smart wristband to the patient, which then connects to the Electronic Patient Care Record (ePCR) device the team carries with them. The ePCR device hosts the identification nanoservice.
    
    
\textit{Phase 2 (on-site) -- Resource discovery and service deployment:} At the site, paramedics start monitoring the patient's pulse rate (PR) and oxygen level ($SpO_2$) with a pulse-oximeter. Later, the team utilizes a portable lightweight Electrocardiogram (ECG) sensor for better observation. 
To get the data from these sensors, nanoservices related to each monitoring task need to be deployed at the ePCR device. For a successful deployment of the nanoservices, the requirements of the nanoservices and the suitable resources must be discovered beforehand.

    
\textit{Phase 3 (in ambulance) -- Resource discovery and service deployment:} At this phase, paramedics move the patient from the site to the ambulance and mount a wireless non-invasive EEG device on the patient's head to monitor and analyze brain activity as the patient becomes unconscious. Furthermore, the monitoring services are connected to the hospital system with the wireless connectivity provided by the ambulance. Moreover, to visualize the patient's status, a visualizer nanoservice needs to be deployed. These new medical resources need to be discovered by the system and deployed at suitable computing nodes in the ambulance, either the ePCR device or another abbulance-mounted monitoring device. 

    
\textit{Phase 4 (from ambulance to hospital) -- Data \& service migration:} When the patient transportation arrives at the hospital, the system must ensure the continuity of the patient monitoring despite the migration of monitoring tasks from ambulance and paramedic team's devices to the devices at the hospital's emergency unit. For this, the system must discover the newly available monitoring devices capable of continuing the patient monitoring and then redeploy the nanoservices to the hospital devices. The use case-related nanoservices are presented in Table~\ref{tab:usecase-services}.

\begin{table}[htbp]
    \centering    
    \caption{Use case nanoservices.}
    \label{tab:usecase-services}
    
    \begin{adjustbox}{max width=0.49\textwidth}
    \begin{tabular}{|m{0.055\textwidth}|m{0.1\textwidth}|m{0.115\textwidth}|m{0.14\textwidth}|} \hline
    
    \rowcolor{gray!30}
    \textbf{Phases} & \textbf{Nanoservices} & \textbf{Resource} & \textbf{Purposes} \\ \hline
    
    Phase 1 & Identification & BLE wristband & Identify the patient. \\  \hline
    
    \multirow{4}{*}{} & ${SpO_2}$ & Oximeter & Detect oxygen saturation level \& Pulse Rate (PR) of the patient. \\ \cline{2-4}
     Phase 2 & Visualizer & Display & Visualize sensor data in a graphical formats \\ \cline{2-4}
     & RabbitMQ & Message broker & Store generated data temporarily. \\ \cline{2-4}
     & ECG & Movesense & Obtain the ECG signal from the patient. \\ \hline
    
    Phase 3 & EEG & \multicolumn{1}{c|}{--} & Collect EEG signal from the patient. \\ \hline
    
    Phase 4 & DB & InfluxDB & Store all sensor data permanently. \\ \hline
    
    \end{tabular}
    \end{adjustbox}
    
    \vspace{-10px}
\end{table}

\section{Three-tier Registry Architecture}
\label{sec:system-model}


\subsection{Proposed System Model}

\subsubsection{Framework setup} 

\begin{table*}[hb]
    \centering
    
    \caption{Registries and their scope of use.}
    \label{tab:registry-scope}
    
    \begin{tabular}{|p{0.21\textwidth}|p{0.15\textwidth}|p{0.15\textwidth}|p{0.15\textwidth}|p{0.21\textwidth}|}
    \hline
    
    \rowcolor{gray!30}
    \textbf{Registry} & \textbf{Location} & \textbf{Purpose} & \textbf{Scope / accessible by} & \textbf{Example Nanoservices} \\ \hline
    
    Local Private Registry (R1) & On-premises (e.g., Ambulance \& Hospital) & Testbed setup & LAN / PAN devices & \hyperref[tab:usecase-services]{Identification (BLE Scanner)}, \hyperref[tab:usecase-services]{Oximeter (${SpO_2}$)}, \hyperref[tab:usecase-services]{Visualizer}, \hyperref[tab:usecase-services]{ECG} \\ \hline
    
    Remote Private Registry (R2) & off-premises (MEC \& cloud e.g., ) & Restricted use (subject to the policy) & Authorized devices & \hyperref[tab:usecase-services]{EEG} \\ \hline
    
    Remote Public Registry (R3) & off-premises (public cloud DockerHub) & Open / generic setup & Any device from anywhere & \hyperref[tab:usecase-services]{InfluxDB}\\ \hline
    
    \end{tabular}
    \vspace{-10px}
\end{table*}

\begin{figure*}[ht]
  \centering
  \includegraphics[trim=1.15in 2.45in 1.30in 2.40in,clip,width=0.85\linewidth]{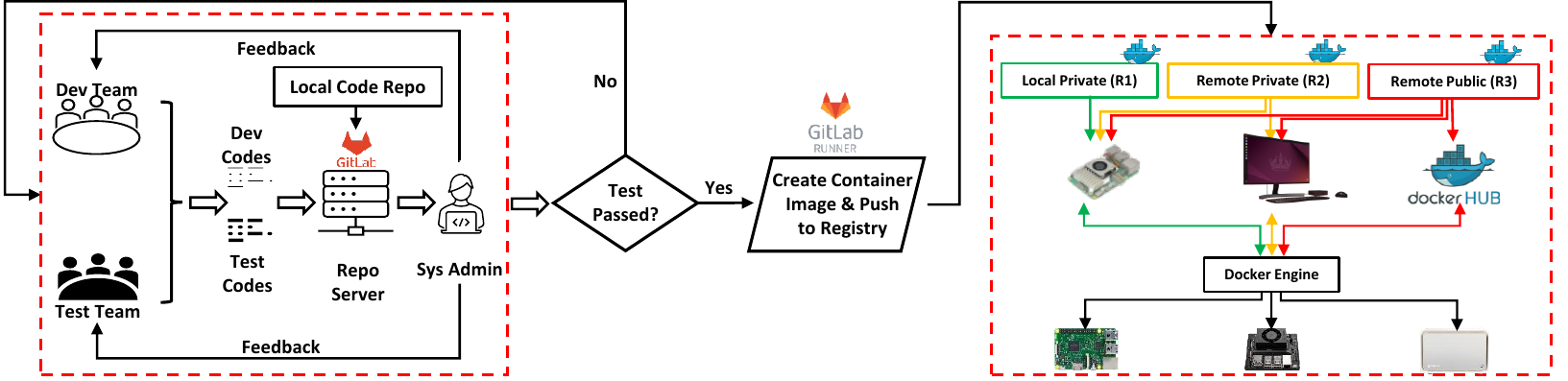}
  \caption{System Model.}
  \label{fig:system-model}
  \vspace{-10px}
\end{figure*} 

\noindent
The success of the deployment
depends on the availability of the underlying network between a computing node and a nanoservice provider, namely a registry server. For the seamless deployment of the nanoservices, we develop a 3-tier registry server to improve reliability, performance, and efficiency across distributed environments. Fig.~\ref{fig:resilient_deployment} represents the system model of the 3-tier registry setup. In the 3-tier registries, the \textit{Local Private Registry (R1)} is inaugurated mainly in the ambulance-mounted devices. Furthermore, the \textit{Remote Private Registry (R2)} is depicted to be configured at the edge computing node of the mobile network, providing connectivity to the accident site and the ambulance, whereas the \textit{Remote Public Registry (R3)} is considered to be deployed on a cloud data center. Table~\ref{tab:registry-scope} shows the intended use and scope of the 3-tier registry setup for resilient deployment.

The service development and deployment model illustrated in Fig. ~\ref{fig:system-model}, where a developer team develops the source code of a nanoservice related to medical sensors and actuators, and a test team develops test code to check related bugs in the nanoservice. The system administrator reviews the code and provides the feedback to the relevant team if needed. The system then generates container images for different computing systems, e.g., for RPi 5, and then uploads those to different registries once the service passes test logic. Here, the R1 provides fast, resilient access to images for on-site workloads; the R2 enables regional caching and distribution, and the R3 serves as the central image repository.

\begin{figure*}[ht]
    \centering
    \includegraphics[trim=0.3in 0.25in 0.25in 0.25in,clip,width=0.75\linewidth]{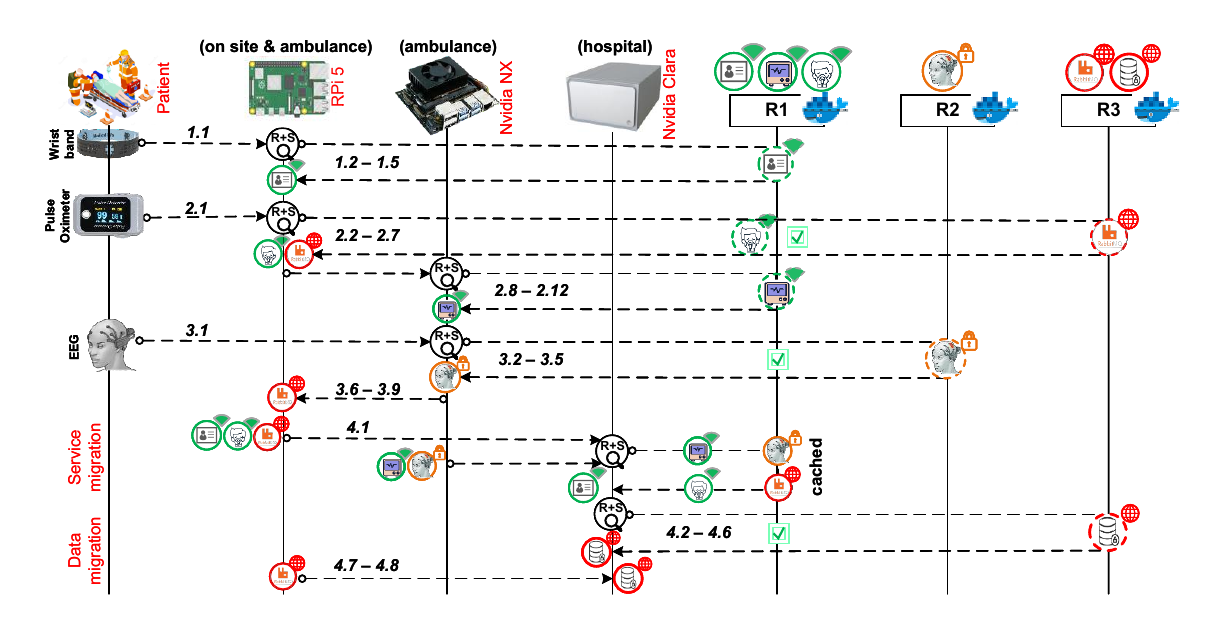}
    \caption{Example sequence of nanoservice deployment in the use case scenario.}
    \label{fig:functional-components-of-sys-model}
    \vspace{-15px}
\end{figure*}

\subsubsection{Functional setup} 
\noindent The proposed system's functionality is illustrated in Fig.~\ref{fig:functional-components-of-sys-model}, where the internal subprocesses of the nanoservices at different phases are denoted by the phase number followed by the subtasks.


  
\textit{Identification:} 
(1.1) When paramedics have attached a BLE-based smart wristband to the patient at the 
site, the wristband starts advertising its BLE ID. (1.2) Upon receiving this advertisement, the resource/service discovery (\textit{R}+\textit{S}) component in a paramedic's ePCR device, depicted with an RPi 5 in our setup, checks if the “BLE Scanner UI” nanoservice is already installed/deployed. (1.3) 
In this case, the nanoservice is not installed, and therefore the \textit{R}+\textit{S} sends a request to the local orchestrator to deploy the “BLE Scanner UI” from a service registry. The orchestror then sends the request to load the service from R1. (1.4) Once loaded, the nanoservice is deployed on the RPi 5. (1.5) Finally, paramedics register the basic patient info at the RPi 5
before starting the sensor data collection.

  
\textit{Oxygen saturation and heart rate monitoring deployment:} 
(2.1) Paramedics configure the pulse oximeter to start monitoring the patient's oxygen saturation ${SpO_2}$ and heart rate \textit{HR} with the patient's wristband ID, which then starts broadcasting with the wristband ID. (2.2) Upon receiving this broadcast, the \textit{R}+\textit{S} checks if the ${SpO_2}$ nanoservice is already deployed at the RPi 5. (2.3) As the \textit{R}+\textit{S} notices that the nanoservice is not installed, it sends a request to the local orchestrator to deploy the nanoservice into the RPi 5. (2.4) Once the ${SpO_2}$ nanoservice is deployed, the RPi 5 starts it. (2.5) Upon a request from ${SpO_2}$ nanoservice to store the data into a \textit{RabbitMQ Broker} nanoservice, the local orchestrator checks if that nanoservice is already deployed at the RPi 5. (2.6) As the \textit{RabbitMQ Broker} is not available in it, the local orchestrator tries to deploy it from R1, which forwards the request to R2, which again forwards the request to R3, as the nanoservice is  available in neither of those. The \textit{RabbitMQ Broker} is found at R3, which returns it to the local orchestrator, which then deploys it. The RPi 5 now starts storing the data into the broker. (2.7) Frequently used nanoservices are cached into R1 and R2 (e.g., \textit{RabbitMQ Broker} in this step) to avoid the excessive deployment delay in the future requests. (2.8) To visualize the sensor data from the broker, the \textit{R}+\textit{S} tries to load the \textit{Visualizer} nanoservice at the Nvidia NX in the ambulance, but fails as it is not available there. (2.9) As a result, the \textit{Visualizer} nanoservice is requested from R1, which happens to have it. (2.10) After the deployment of \textit{Visualizer} it starts receiving ${SpO_2}$ and \textit{HR} data from RPi 5, (2.11) once those are published. (2.12) Finally, the \textit{Visualizer} nanoservice starts visualizing the ${SpO_2}$ and \textit{HR} data in a display.

  
\textit{Brain monitoring deployment:} 
(3.1) In the ambulance, the paramedics configure the EEG device to start monitoring the patient's brain status similarly to the previous case. The EEG sensor device is then registered to the system with the patient ID. (3.2) The \textit{R}+\textit{S} at the Nvidia NX (at ambulance) requests the \textit{EEG} nanoservice. (3.3) As it is not installed, the local orchestrator tries to deploy the nanoservice from R1, which forwards the request to R2. (3.4) The \textit{EEG} nanoservice is deployed from R2 (and cached into R1). (3.5) After that, the Nvidia NX starts \textit{EEG}. (3.6) Once initiated, the EEG sensor start capturing and publishing EEG data to the Nvidia NX. (3.7) The RabbitMQ Broker now subscribes to EEG data from the Nvidia NX. Finally, RPi 5 acquires (3.8) and visualizes (3.8) the EEG data in a in a display.

  
\textit{Service and data migration:} 
(4.1) As soon as the ambulance reaches the hospital, \textit{R}+\textit{S} at the RPi 5 initiates the migration of the monitoring services to Nvidia Clara (depicting the hosting node at the hospital), by instructing it to deploy the needed nanoservices (\textit{BLE Scanner}, \textit{RabbitMQ Broker}, \textit{${SpO_2}$}, \textit{EEG}). Similarly to the previous phases, the \textit{R}+\textit{S} initiates the deployment of these nanoservices. As all of these recently used nanoservices have been cached by R1 during the previous deployments, those are deployed from there. (4.2) To migrate the previously acquired patient data, the \textit{R}+\textit{S} at the Nvidia Clara requests the deployment of the \textit{DB} nanoservice. After discovering that it is not locally installed, (4.3) the local orchestrator first tries to deploy it from R1 and then (4.4) from R2, and finally (4.5) from R3, from where it is successfully deployed. (4.6) After the successful deployment, the Nvidia Clara starts the \textit{DB} nanoservice. (4.7) The Nvidia Clara subscribes to the RPi 5 and Nvidia Nx to transfer the relevant sensor data from those. (4.8) Finally, RPi 5 and Nvidia Nx publish all requested data to the Nvidia Clara. Once finished, the monitoring can continue at the hospital premises. The nanoservices no longer used at RPi 5 and Nvidia Nx can either remain deployed or be undeployed based on the need.
  

\begin{figure*}[ht!]
\centering

\begin{minipage}{0.33\textwidth}
    \vspace{3px}
    \centering
    \includegraphics[trim=0.0in 0.0in 0.0in 0.65in, clip, height=0.70\linewidth, width=\linewidth]{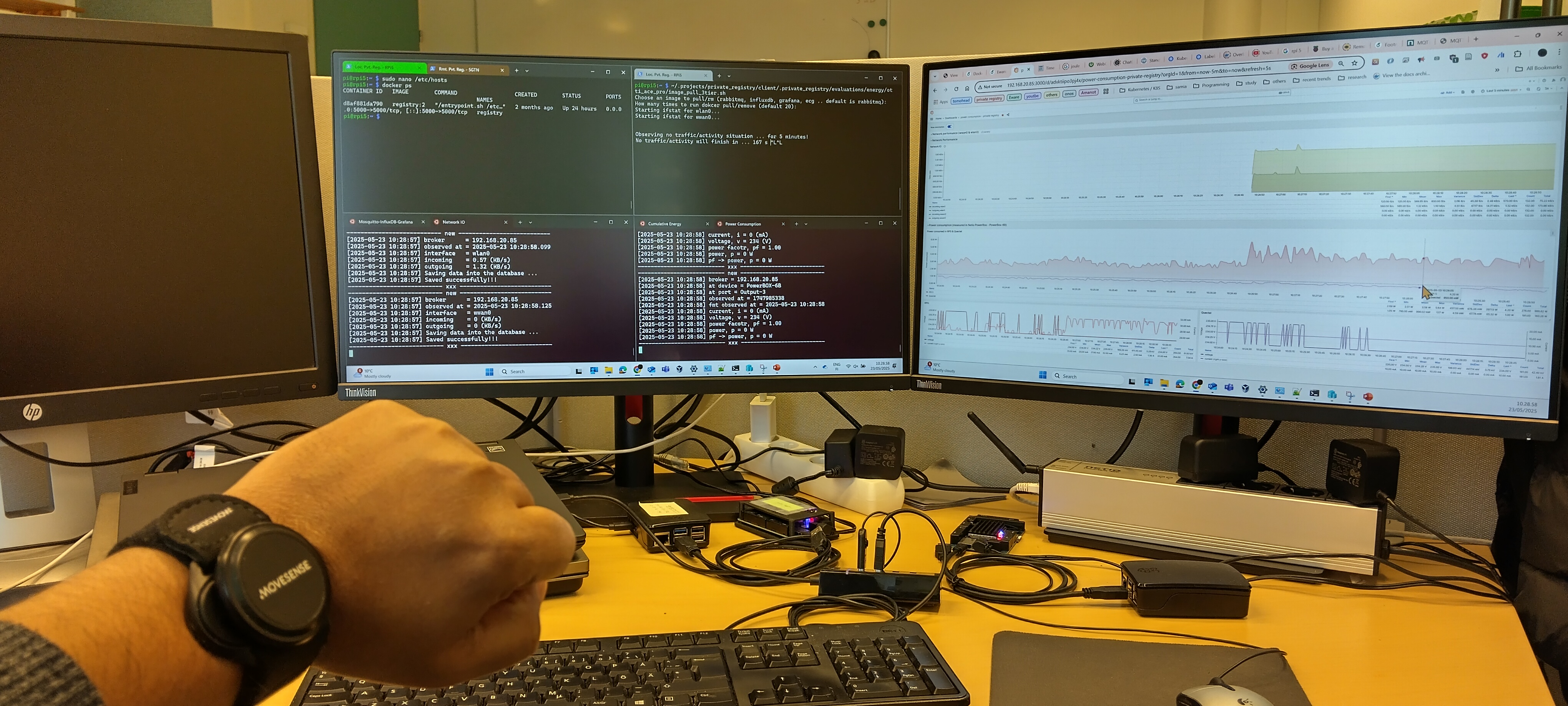}
    \captionof{figure}{Implementation setup.}
    \label{fig:testbed}
\end{minipage}
\hfill
\begin{minipage}{0.66\textwidth}
    \centering
    \vspace{-3px}
    \includegraphics[trim=0.3in .3in .3in .3in, clip, width=\linewidth]{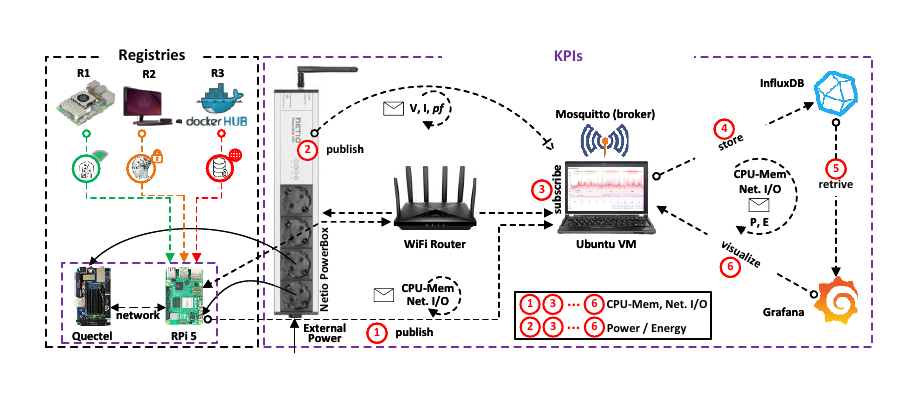}
    \captionof{figure}{KPIs evaluation at RPi 5 (during $\rightarrow$ no traffic and pulling images).}
    \label{fig:evaluation-setup}
\end{minipage} 
\vspace{-10px}
\end{figure*}

\subsection{Prototype Implementation} 
\label{sec:implementation}

\noindent Nanoservices are deployed into worker nodes from the different registries (R1, R2 \& R3), based on the usage and scope as defined in Table \ref{tab:registry-scope}. In our 3-tier registry setup, the \textit{Identification}, ${SpO_2}$, \textit{ECG} nanoservices (required by RPi 5), and \textit{Visualizer} (required by 
Nvidia NX) are primarily deployed from the local private registry (R1). The \textit{EEG} nanoservice is mainly deployed from the 5GTN-based remote private registry (R2). Additionally, all other publicly available Docker images, for instance, the \textit{RabbitMQ}, \textit{DB}, etc., are deployed directly from the Docker Hub public registry (R3). In addition, R1 and R2 keep the services deployed from upper levels in their caches for a predefined time, in order to serve faster the frequently deployed upper-level services.
Figs.~\ref{fig:testbed} and ~\ref{fig:evaluation-setup} depict the testbed setup of the 3-tier registry servers.

\textit{\textbf{Hardware specifications:}} 
The R1 registry server is deployed on a Debian GNU/Linux 12 (bookworm) OS-based RPi 5 having a Cortex-A76 ARM (aarch64) CPU along with 8 GB memory (RAM) and 32 GB storage. A 5GTN-based VM (in the University of Oulu) is used for the R2 registry server with Ubuntu 24.04.2 LTS (Noble Numbat) having an Intel(R) Xeon(R) Gold 6230 CPU along with 8 GB memory (RAM) and 64 GB storage. The node hosting the R3 DockerHub depends on the data center hardware. 

    
\textit{\textbf{Software specifications:}}
The nanoservices are virtualized as Docker containers. To deploy the containerized nanoservices, Docker Engine 27.3.1 is used in each node.
Containerized nanoservices are stored in R1, R2 and R3 registry servers. To build the private registries R1 and R2, \textit{registry:2} Docker image is used. Furthermore, \textit{portainer/portainer-ce:2.21.4} Docker image is used optionally as the registry UI. 
Finally, GitLab 17.4.1-ee.0 and GitLab Runner v17.4.0 are used to build the entire continuous integration and continuous deployment (CI/CD) pipelines as depicted in Fig.~\ref{fig:system-model}. 



\section{Evaluation Setup}
\label{sec:evaluations}


\noindent 

\noindent The performance of the retrieval and deployment of Docker images in our scenario---and in general---depends on the physical and logical distance between the worker node and the resource registry. Furthermore, there are significant differences in the resource consumption profiles---including CPU, network, storage, and energy resources---between the deployment from different registries. To evaluate the feasibility of the proposed approach, we conducted a Proof-of-Concept (PoC) experiment and identified a set of Key Performance Indicators (KPI)s to measure the performance and efficiency of the deployment. The evaluation setup used in the experiment is presented in Fig.~\ref{fig:evaluation-setup}. 
In the experiment, nanoservice containers (as described in Table~\ref{tab:usecase-services}) are downloaded and deployed to a local RPi 5 device, acting as the registry client, from three different registries: R1, R2, and R3. A Cudy 5G NR AX3000 WiFi 6 router and a Quectel RM500Q-GL 5G modem enable the RPi 5 to download the required nanoservices from the registry servers through Wi-Fi and 5G network interfaces, respectively. The KPIs are monitored separately on the RPi 5 and the Netio PowerBox, as discussed below. 

\noindent
\subsection{KPI definitions:}


            
\textit{Performance:} We measured the performance of the deployment by observing the round-trip time (RTT) in milliseconds (ms) and the number of hops (\#) between the RPi 5 and registry servers and the overall download and deployment time in seconds (s) of deploying the used nanoservices from each registry service. We considered two different scenarios---cached and non-cached---depicting the scenarios where the registry server path is already known or unknown to the RPi 5.
    
\textit{Computational load:} The KPIs related to computational load (CPU and storage) were observed directly in RPi 5. The CPU load is presented in percentage (\%) which is basically the average number of the processes that are either (i) running on the CPU or (ii) waiting in the runnable queue while the CPU is busy. 
The average storage utilization of images was observed in megabytes (MB) during the deployment. Both observations were carried out with Linux OS's resource-monitoring tools. 

    
\textit{Network load:} Similarly, the inflicted network load, i.e., incoming/outgoing network traffic, was observed at the RPi 5 registry client while a nanoservice container was being downloaded from different registries. The network load was measured in Megabytes per second (MB/s) with Linux OS's resource monitoring tools.

    
\textit{Power consumption:} We used Netio PowerBox 4KF to acquire the power consumed by the RPi 5 and the Quectel modem. At first, the RPi 5 and the Quectel modem were mounted on 2 physical sockets in Netio, and then turned on through the Netio dashboard. Later, the voltage in (\textit{Volts, V}), the current I in (\textit{Ampere, A}) and the true power factor (\textit{pf}) were obtained from Netio, while the Docker images were being downloaded from different registries.





\begin{algorithm}[H]
\caption{KPI analysis for 3-tier registry}
\label{algo:3-tier-registry-evaluation}
\begin{algorithmic}[1]
\State Registries, $R = \{R1, R2, R3\}$
\State Users, $U = \{user1, user2, user3, ..., user_n\}$
\State Tags, $T = \{t1, t2, t3, ..., t_n\}$
\State \textbf{Inputs:} $\leftarrow image, iterations$
\State // Start: KPI monitoring processes (with MQTT)
\For{ $(registry, user, tag) \in (R, U, T)$ }
    \State sleep(5*60)\ \ \ \ // no-traffic -- for 5 minutes 
    \For{$i = 1 \leftarrow iterations$}
        \State $sleep(20)$
        \State $docker\ pull\ registry/user/image:tag$
        \State $sleep(10)$
        \State $docker\ rmi\ registry/user/image:tag$
    \EndFor
\EndFor
\State sleep(5*60)\ \ \ \ // no-traffic -- for 5 minutes
\State // Stop: KPI monitoring processes (with MQTT)
\end{algorithmic}
\end{algorithm}


\noindent
\subsection{KPI data collection:}

\noindent
The KPIs for the 3-tier registry setup are collected with a Linux bash script while a Docker image (i.e., nanoservice) is downloaded to RPi 5 from the registries, as shown in Algorithm~\ref{algo:3-tier-registry-evaluation}. The script accepts two inputs: (1) the name of the nanoservice image and (2) the number of iterations. 
With MQTT, the script starts monitoring the KPIs for 5 minutes for the idle period before downloading a Docker image from any registry. Generally, a Docker image is not downloaded if it is already in the local storage in RPi 5. To see the impact on the KPIs during downloading a nanoservice, the related Docker image is deleted from the local storage each time. In every iteration, there is a 20s pause before downloading and a 10s pause before deleting the image. 
At the end, the script terminates all MQTT-based KPI monitoring processes.

\vspace{5px}
\noindent
\subsection{KPI visualization:}
\vspace{3px}

\noindent
In the experiment, a WSL-based Ubuntu 24.04 VM
was used to store and visualize the measured data.
Fig.~\ref{fig:evaluation-setup} presents the subtasks with red circles: (1) At first, the CPU, storage, and network utilization are acquired directly from RPi 5 using Linux tools; (2) the voltage (V), current (I), and power factor (pf) values are acquired from Netio and then 
sent to the Ubuntu VM via WiFi by using the MQTT-based pub-sub protocol; (3) the RPi 5 and Netio publish the acquired data to the Mosquitto 2.0.15 MQTT broker; finally, (4) the inflicted computational and network load data are stored in InfluxDB 2.7.10 as soon as the data are available in the Ubuntu VM. Equation \ref{eqn:power} is used to calculate the instantaneous power (watt, W) 
with Netio-acquired \textit{V}, \textit{I} and \textit{pf}.
During the deployment of a nanoservice, equation \ref{eqn:energy} is used to calculate the energy consumption with the mean power ($\overline{P}$) and mean download time ($\Delta t$) observed for a nanoservice. 

\begin{equation}
    \label{eqn:power}
    Power, P = V \times I \times pf
\end{equation}

\vspace{-20px}
\begin{equation}
    \label{eqn:energy}
    Energy, E = \overline{P} \times \Delta t
\end{equation}

Both the power and energy consumptions are observed before storing in InfluxDB. (5 \& 6) Finally, Grafana 10.3.1 is used to visualize all these KPIs in a graphical format in a web browser through a dashboard.

\section{Results and Analysis}

\subsection{Performance}

\noindent 
We evaluate the performance of service deployments while different nanoservices are being downloaded from registries R1, R2, and R3. The download event is basically divided into two parts, e.g., registry discovery and actual content, or the nanoservice download. Thus, at first, the registry discovery is observed with the round-trip time (RTT) and the number of hops between the client (RPi 5) and different registry servers. 
Here, the network hops indicate the links between the registry client, i.e., RPi 5, and the target registry server, including the intermediate networking devices (routers, etc.).  
%
%
%
According to the observation, only one hop is required to reach the R1 
registry, 
as it is deployed locally. The mean latency in this case is 46.9 ms. To catch up to the R2 
registry, 
5-8 hops are needed in our case, resulting in 186.24 ms mean latency. The R3 is reached by 28-29 hops, resulting in the mean latency of 1435.18 ms.



    
       
      
      
      

\vspace{-10px}
\begin{figure}[h!]
  \centering
  \includegraphics[trim=0.16in 0.13in 0.20in 0.15in, clip, width=1.0\linewidth]{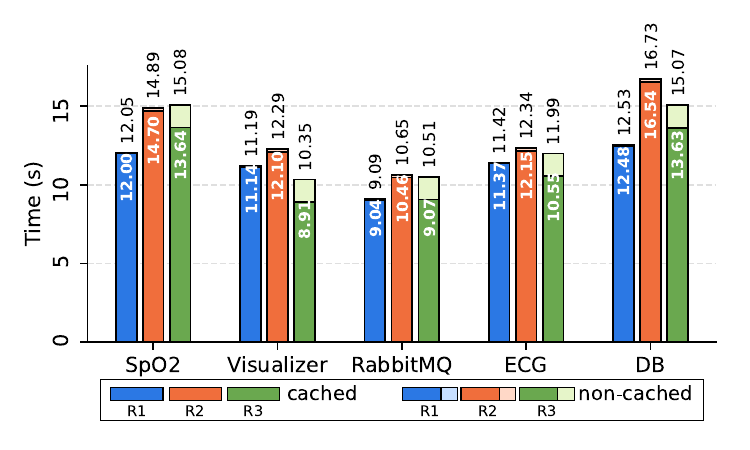}
  \caption{Nanoservices download time for R1, R2 \& R3.}
  \label{fig:setup_delay}
  \vspace{-10px}
\end{figure}

In the next phase, we observed the overall time consumed by the deployment sequences for the five nanoservices used in our scenarios, when deployed from R1, R2 and R3. These results are presented in Fig.~\ref{fig:setup_delay}. For this, we consider two different scenarios: cached and non-cached. In real life, this scenario depends on whether the registry server path is already known or unknown to the RPi 5. The summary of the observations is presented as stacked 
blue, orange, and green bars 
while the nanoservices are downloaded from R1, R2 and R3, respectively. The lower part of the stacked bars represent the download time when the routing path is known (cached) to RPi 5, while the upper part indicates the registry discovery time when the routing path is not known or cached at RPi 5. Therefore, in non-cached scenarios, the download time is the sum of registry discovery and actual download time. For instance, the RPi 5 takes 12.00 s to download the ${SpO_2}$ nanoservice from R1 when the routing path is known to RPi 5, and another 0.05 s is added if the registry path to R1 needs to be discovered. Thus, it takes 12.05 s in total to download the ${SpO_2}$ nanoservice while the registry path is not known beforehand. The RPi 5 downloads the same nanoservice from R2 in 14.70 (cached) / 15.89 s (non-cached), and from R3 in 13.64 s (cached) / 15.08 s (non-cached). The observations suggest that the RPi 5 reliably downloads a nanoservice from R1 when the external network is not in function, as predicted. When the same nanoservice is downloaded from R2 and R3, slightly less time is needed when the nanoservice is downloaded from R3 rather than R2, as R3 is maintained through a content delivery network (CDN), which makes it more optimized apart from its other higher-capacity physical configurations. The results for the other nanoservices (Visualizer, RabbitMQ, ECG and DB) can also be read from Fig.~\ref{fig:setup_delay}, showing similar trends. 

\begin{table*}[ht!]
    \centering
    \caption{CPU Load 
    observed in RPi 5 while nanoservices are downloading from R1, R2 \& R3.}
    \label{tab:cpu-mem-load}
    \begin{adjustbox}{max width=\textwidth}
    \begin{tabular}{|l|c|c|c|c|c|c|c|c|}
    
        \hline
        \rowcolor{gray!30}
        & \multicolumn{8}{c|}{\textbf{Last 15 minutes CPU load (in percentage, \%) at RPi 5}} \\ \cline{2-9}
        
        \rowcolor{gray!30}
        & \multicolumn{2}{c|}{\textbf{No traffic}} & \multicolumn{2}{c|}{\textbf{Local Private Registry (R1)}} & \multicolumn{2}{c|}{\textbf{Remote Private Registry (R2)}} & \multicolumn{2}{c|}{\textbf{Remote Public Registry (R3)}} \\ \cline{2-9}
        \rowcolor{gray!30}
        \multirow{-3}{*}{\textbf{Nanoservices}} & 
        \textbf{Mean 
        } & \textbf{Std. Dev. 
        } & 
        \textbf{Mean 
        } & \textbf{Std. Dev. 
        } & 
        \textbf{Mean 
        } & \textbf{Std. Dev. 
        } & 
        \textbf{Mean 
        } & \textbf{Std. Dev. 
        } \\
        \hline
        
        ${SpO_2}$ &  &  & 
        \cellcolor[HTML]{fffabc} 20.0 & 2.5 & 
        \cellcolor[HTML]{fffabc} 22.5 & 2.5 & 
        \cellcolor[HTML]{fffabc} 25.0 & 2.5 \\
        \cline{1-1} \cline{4-9}
        
        Visualizer &  &  & 
        \cellcolor[HTML]{affaaa} 15.0 & 2.5 & 
        \cellcolor[HTML]{affaaa} 27.5 & 5.0 & 
        \cellcolor[HTML]{affaaa} 27.5 & 2.5 \\
        \cline{1-1} \cline{4-9}
        
        RabbitMQ &  &  & 
        \cellcolor[HTML]{fbdaff} 15.0 & 0.0 & 
        \cellcolor[HTML]{fbdaff} 20.0 & 2.5 & 
        \cellcolor[HTML]{fbdaff} 17.5 & 0.0 \\
        \cline{1-1} \cline{4-9}
        
        ECG &  &  & 
        \cellcolor[HTML]{a0c5e5} 12.5 & 2.5 & 
        \cellcolor[HTML]{a0c5e5} 15.0 & 2.5 & 
        \cellcolor[HTML]{a0c5e5} 22.5 & 2.5 \\
        \cline{1-1} \cline{4-9}
        
        DB & \multirow{-5}{*}{10.0} & \multirow{-5}{*}{0.0} &
        \cellcolor[HTML]{d3d3a3} 27.5 & 2.5 & 
        \cellcolor[HTML]{d3d3a3} 22.5 & 0.0 & 
        \cellcolor[HTML]{d3d3a3} 30.0 & 5.0 \\
        \hline
    \end{tabular}
    \end{adjustbox}
    \vspace{-10px}
\end{table*}

\subsection{Computational efficiency}

\noindent We evaluate the computational efficiency by two measurements, the inflicted computational load, and the inflicted storage consumption. The results of these measurements are presented in the following subsections. 

\subsubsection{\textbf{Computational load}}

\begin{figure}[h!]
    \centering
    \includegraphics[width=1.0\linewidth]{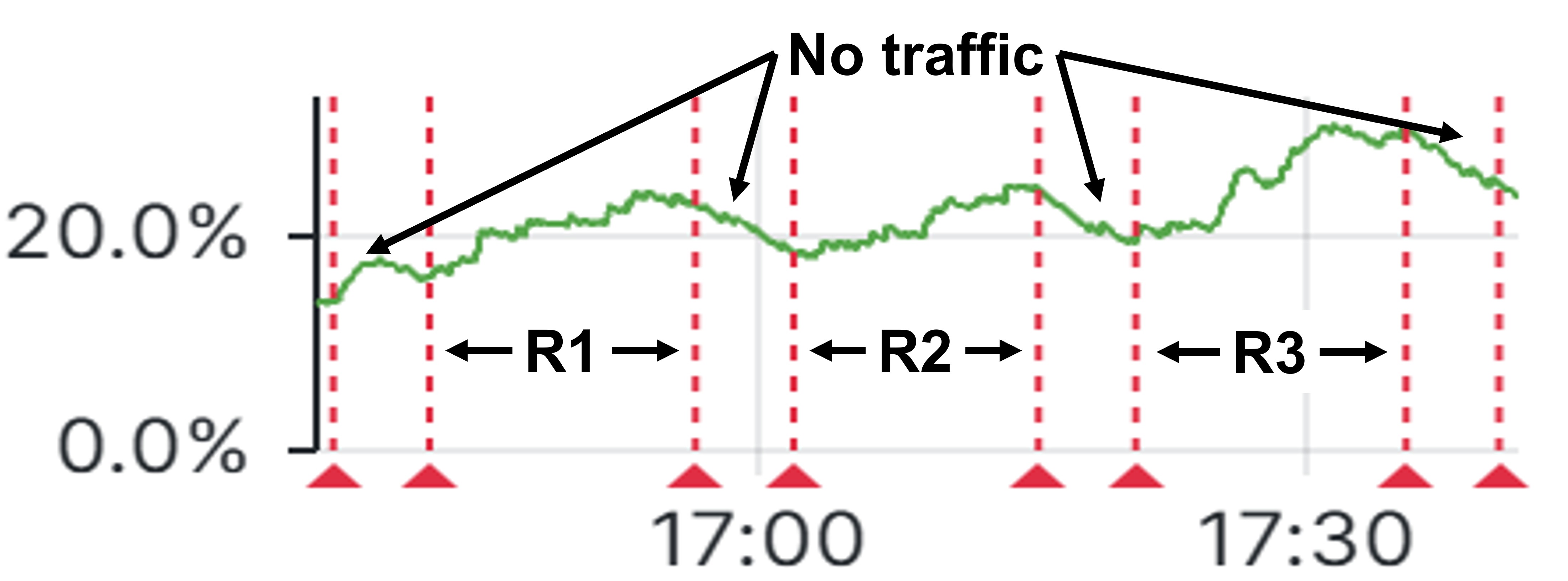}
    \caption{Inflicted computational load at RPi 5.}
    \label{fig:cpu_mem_usage}
    \vspace{-10px}
\end{figure}

We use the CPU load to evaluate the inflicted computational load on the RPi 5 registry client during downloading nanoservices from different registries. 
Fig. \ref{fig:cpu_mem_usage} shows the CPU load in RPi 5 during the $SpO_2$ nanoservice download from different registries, and the no-traffic situations in between. 

In our evaluation scenario, CPU load is initially observed for 5 minutes with no deployment activity in order to observe the baseline CPU load. Then, CPU load is observed for 20 subsequent nanoservice downloads from different registries. With almost all the nanoservices, all 20 downloads from 
any registry
were accomplished roughly in 15 minutes. Thus, a 15 minute CPU load is observed for a nanoservice while it is downloading from a registry. The overall result is consolidated in Table~\ref{tab:cpu-mem-load}.

During the no-traffic situation, the mean CPU load at the RPi 5 registry client was measured to be around 10\%. While the $SpO_2$ nanoservice was downloaded from the R1, R2, and R3 registry servers, the average CPU load reached 20\%, 22.5\%, and 25\%, respectively. In other words, the CPU load at the RPi 5 were increased by 10\%, 12.5\%, and 15\%. In the case of visualizer nanoservice, R1, R2, and R3 increased the CPU load at RPi 5 by 5\%, 17.5\%, and 17.5\%, respectively. In the case of RabbitMQ, the respective increases were 5\%, 10\%, and 7.5\%, while in the case of ECG nanoservice, the readings were 2.5\%, 5\%, and 12.5\%, and 17.5\%, 12.5\%, and 20\% in the case of the DB nanoservice. \\

\subsubsection{\textbf{Storage consumption}}



Different components are required for building the virtualized system and CI/CD pipelines to deploy the nanoservices. The step-by-step storage consumption is described below. 

Docker Engine requires five different packages.
Table~\ref{tab:storage_consumption} (columns 1-4)
shows the storage consumed by these packages on RPi 5 and Ubuntu VM-based 5GTN to set up the R1 and R2 registry servers. 
In RPi 5, the \textit{docker-ce-cli}, \textit{docker-ce}, \textit{containerd.io}, \textit{docker-buildx-plugin}, and \textit{docker-compose-plugin} consume 40.1, 79.4, 97.5, 80.1, and 62.2 MB, respectively. On the other hand, in 5GTN VM, these packages consume 41.5, 111, 121, 82.5, and 53.9 MB, respectively. \textit{Docker-buildx-plugin} and \textit{docker-compose-plugin} are optional, although helpful for debugging and deploying the Docker containers in a node. All Swarm nodes (both workers and managers) require \textit{docker-ce} and \textit{containerd.io} packages. Additionally, each Swarm manager requires \textit{docker-ce-cli}. In our setup, each RPi 5-based Swarm worker and manager consumes 176.9 MB and 217 MB, respectively. Furthermore, each Ubuntu VM-based Swarm worker and manager consumes 232 MB and 273.5 MB, respectively.


Table~\ref{tab:storage_consumption} (columns 5-8)
shows the storage consumption for setting up the CI/CD pipeline. 
In the experiment, 
GitLab consumes 3.45 GB, while GitLab Runner consumes 798 MB of disk storage at the Ubuntu-based VM. 
The \textit{registry:2} Docker image
consumes 25 MB and 25.4 MB at the RPi 5 and the Ubuntu-based VM, respectively, which are used to deploy R1 and R2. The \textit{portainer/portainer-ce:2.21.4} Docker image consumes 302 MB in the Ubuntu-based VM, which is used for the registry UI. 

Docker internally compresses the containerized nanoservices 
before storing those into registry servers. Therefore, they need to be decompressed at the local node after downloading from a Swarm worker. Table~\ref{tab:storage_consumption} (columns 9-12)
presents the storage consumed by use case nanoservices at different registries (compressed) and on RPi 5's local disk (decompressed). At the registry servers, the $SpO_2$, Visualization, RabbitMQ, ECG, and DB approximately consume 128.56, 100.46, 58.21, 101.45, and 155.24 MB. At the RPi 5, the decompressed nanoservices approximately consume 365, 399, 134, 215, and 416 MB, respectively. 

\begin{table*}[ht!]
    \centering
    \caption{Storage consumption.}
    \label{tab:storage_consumption}

    ${\colorboxrule[mygreen]{10pt}{5pt} + \colorboxrule{10pt}{5pt} \Rightarrow managers \hspace{0.45cm}}$
        ${\colorboxrule{10pt}{5pt} \Rightarrow workers \hspace{0.45cm}}$
        ${\colorboxrule[orange!35]{10pt}{5pt} \Rightarrow optional 
        }$
    
    \begin{adjustbox}{max width=\textwidth}
    \begin{tabular}{| l | c | c | c | c | c | c | c | c | c | c | c |}
      \hline 

      \rowcolor{gray!30}
      \multicolumn{4}{|c|}{\textbf{for infrastructure}} & \multicolumn{4}{c|}{\textbf{for CI/CD pipeline}} & \multicolumn{4}{c|}{\textbf{for use case}} \\ \hline 
      
      \rowcolor{gray!30}              
       &  & \multicolumn{2}{c|}{\textbf{Size (MB) at}} 
       &  &  & \multicolumn{2}{c|}{\textbf{Size}} 
       &  &  & \multicolumn{2}{c|}{\textbf{Size (MB) at}} \\ \cline{3-4} \cline{7-8} \cline{11-12} 
      
      \rowcolor{gray!30}        
      \multirow{-2}{*}{\textbf{Packages}} & \multirow{-2}{*}{\textbf{Version}} & \textbf{RPi 5} & \textbf{5GTN VM} 
      & \multirow{-2}{*}{\textbf{Packages}} & \multirow{-2}{*}{\textbf{Version}} & \textbf{RPi 5} & \textbf{5GTN VM} 
      & \multirow{-2}{*}{\textbf{Nanoservices}} & \multirow{-2}{*}{\textbf{Tags}} & \textbf{server\textsuperscript{3}} & \textbf{client\textsuperscript{4}} \\ \hline

      \cellcolor[HTML]{d9ffd9} docker-ce-cli   & \cellcolor[HTML]{d9ffd9} 27.3.1  & \cellcolor[HTML]{d9ffd9} 40.1  & \cellcolor[HTML]{d9ffd9} 41.5
      & Gitlab & 17.4.1-ee.0  & -- & 3.45 GB 
      & ${SpO_2}$ & v2 & 128.56 & 362 \\ \hline

      \cellcolor[HTML]{ccd9ff} docker-ce & \cellcolor[HTML]{ccd9ff} 27.3.1  & \cellcolor[HTML]{ccd9ff} 79.4  & \cellcolor[HTML]{ccd9ff} 111 
      & Gitlab Runner & v17.4.0  & -- & 798 MB
      & Visualization & 10.3.1 & 100.46 & 399 \\ \hline

      \cellcolor[HTML]{ccd9ff} containerd.io & \cellcolor[HTML]{ccd9ff} 1.7.22 & \cellcolor[HTML]{ccd9ff} 97.5 & \cellcolor[HTML]{ccd9ff} 121
      & Registry & 2 & 25 MB & 25.4 MB
      & RabbitMQ & 3.12.6-alpine & 58.21 & 134\\ \hline

      \cellcolor[HTML]{ffd2a6} docker-buildx-plugin\textsuperscript{1} & \cellcolor[HTML]{ffd2a6} 0.17.1 & \cellcolor[HTML]{ffd2a6} 80.1 & \cellcolor[HTML]{ffd2a6} 82.5
      & \cellcolor[HTML]{ffd2a6} Portainer & \cellcolor[HTML]{ffd2a6} & \cellcolor[HTML]{ffd2a6} & \cellcolor[HTML]{ffd2a6}
      & ECG & v1 & 101.45 & 215 \\ \cline{1-4} \cline{9-12}

      \cellcolor[HTML]{ffd2a6} docker-compose-plugin\textsuperscript{2} & \cellcolor[HTML]{ffd2a6} 2.29.7 & \cellcolor[HTML]{ffd2a6} 62.2 & \cellcolor[HTML]{ffd2a6} 63.9
      & \cellcolor[HTML]{ffd2a6} (Registry UI) & \cellcolor[HTML]{ffd2a6} \multirow{-2}{*}{2.21.4} & \cellcolor[HTML]{ffd2a6} \multirow{-2}{*}{--} & \cellcolor[HTML]{ffd2a6} \multirow{-2}{*}{302 MB}
      & DB & 2.7.10 & 155.24 & 416 \\ \hline
      
    \end{tabular}
    \end{adjustbox}

      \vspace{1px}
      1 $\rightarrow$ for building cross-platform images, 
      \hspace{25px} 2 $\rightarrow$ 
      for deploying multi-containers, 
      \hspace{25px} 3 $\rightarrow$ compressed, 
      \hspace{25px} 4 $\rightarrow$ decompressed.

      \vspace{-10px}
\end{table*}



\subsection{Network efficiency}

\begin{figure}[h!]
    \centering
    \includegraphics[trim=0.16in 0.18in 0.28in 0.15in, clip, width=1.0\linewidth]{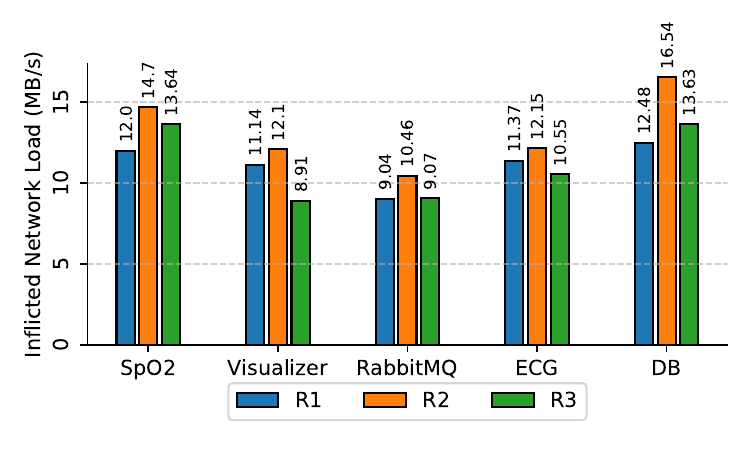}
    \caption{Inflicted network load.}
    \label{fig:net_util_spo2}
    \vspace{-5px}
\end{figure}

\noindent The efficient use of network resources is an important factor of resource-efficient service deployment. For this, we compare the total local network load inflicted by deploying the nanoservices from alternative registries: R1, R2 and R3. In our scenario, the R1 resides within the local WiFi network, and therefore the RPi 5's built-in WiFi is used for nanoservice downloads from the R1. Furthermore, the remote registries are connected through a 5G network connectivity. Therefore, a Quectel modem is used for deploying nanoservices from remote registries R2 and R3. Fig. \ref{fig:net_util_spo2} shows the average network load 
while the $SpO_2$ nanoservice was downloading from registries R1, R2, and R3.

According to the experiment, the RPi 5 downloaded the $SpO_2$ nanoservice at 12 MB/s, 14.7 MB/s, and 13.64 MB/s from R1, R2, and R3. The visualizer nanoservice is downloaded at the RPi 5 from R1, R2, and R3 at 11.14 MB/s, 12.10 MB/s, and 8.91 MB/s, respectively. Furthermore, the RabbitMQ nanoservice is downloaded at 9.04 MB/s, 10.46 MB/s, and 9.07 MB/s. The respective results for the ECG nanoservice were 11.37 MB/s, 12.15 MB/s, 10.55 MB/s, 12.48 MB/s, 16.54 MB/s and 13.63 MB/s for the DB nanoservice.
 
According to the results, in most cases the RPi 5 utilizes the least bandwidth when the nanoservices are downloaded from R1. In the case of $Visualizer$ and $ECG$ nanoservices, more bandwidth is utilized while  downloading from R1 rather than R3. However, the standard deviations ($\sigma_R1$), and ($\sigma_R3$), for the transfers from R1 are lower, indicating higher stability of the end-to-end network path. The highest network load is inflicted while the nanoservices are downloaded from R2. The commercial registry server R3 performs better than our on-premises registry server R2, as the R3 has more computational capacity than the R2. However, R2 is more stable than R3 when the standard deviation is considered. \\

\subsection{Energy-efficiency}

\begin{figure}[ht!]
  \centering
  \includegraphics[trim=0.0in 0.0in 0.0in 0.0in, clip, width=1.0\linewidth, width=\linewidth]{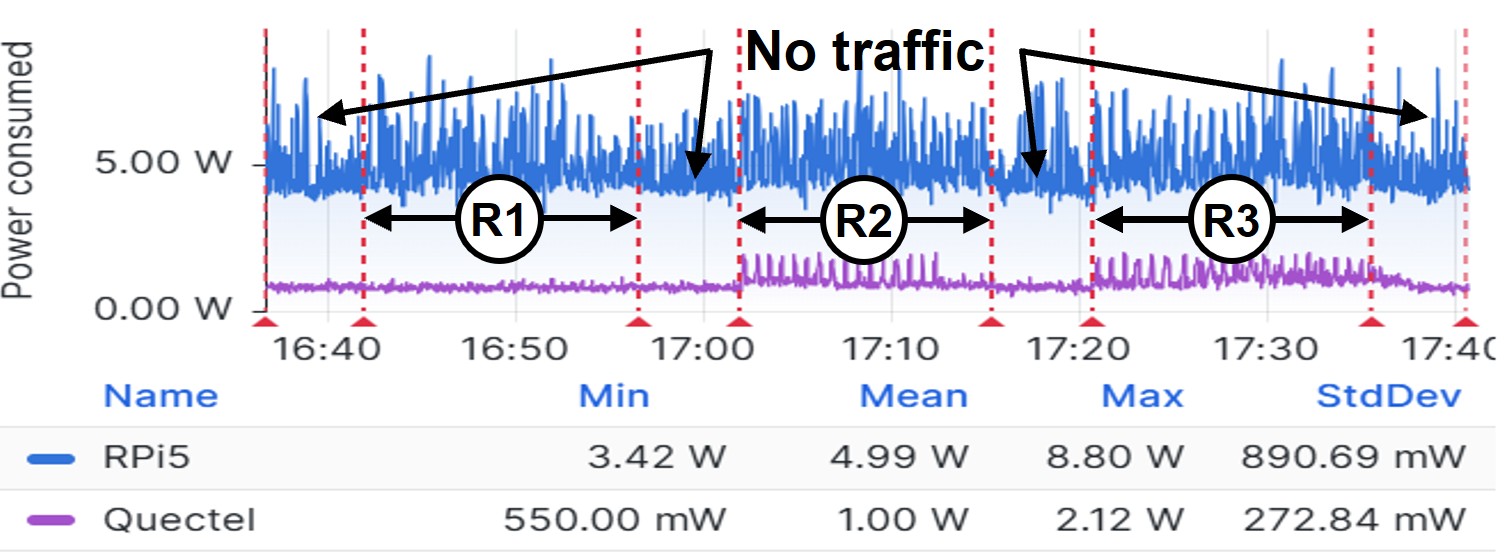}
  \caption{Mean power consumption of RPi 5 \& Quectel.}
  \label{fig:time-vs-power}
\end{figure}

\noindent We also observed energy consumption of the RPi 5 (device energy consumption excluding the networking energy consumption) and the Quectel modem (the networking energy consumption) while the nanoservices are downloaded from the R1, R2, and R3.
Table~\ref{tab:service-wise-energy-consumption} presents the observed mean power consumption (${\overline{P}}$), mean download time (${\Delta t}$) and the calculated mean energy consumption ($E$) for a deployment event of each nanoservice. To observe the baseline power consumption, the mean power consumption was observed for 5 minutes with no deployment activity before, in between and after the deployments. To exemplify the measurement process, Fig.~\ref{fig:time-vs-power} shows the real-time power consumption for the ${SpO_2}$ nanoservice when it downloads from R1, R2, and R3 registries.

\begin{table*}[ht!]
    \centering
    \caption{Energy consumption at RPi 5 \& Quectel while images are downloading from R1, R2, \& R3.}
    \label{tab:service-wise-energy-consumption}
    \begin{adjustbox}{max width=\textwidth}
    \begin{tabular}{| l | 
    c | c | c | c | c | c | c | c 
    | c | c | c | c | c | c | c | c | c |}
      \hline 

      \rowcolor{gray!30}
       & \multicolumn{8}{c|}{\textbf{Mean power consumption (${\overline{P}}$ in W) of }} 
       & \multicolumn{9}{c|}{\textbf{Mean energy consumption ($E$ in J) for a single download event}} 
       \\ \cline{2-18} 

      \rowcolor{gray!30}
      & \multicolumn{4}{c|}{\textbf{RPi 5}} & \multicolumn{4}{c|}{\textbf{Quectel}} 
      & \multicolumn{3}{c|}{\textbf{RPi 5}} & \multicolumn{3}{c|}{\textbf{Quectel}} & \multicolumn{3}{c|}{\textbf{Total (RPi 5 + Quectel)}} \\ \cline{2-18}
      
      \rowcolor{gray!30}
      \multirow{-3}{*}{\textbf{Nanoservices}} & \textbf{NT\textsuperscript{*}} & \textbf{R1\textsuperscript{**}} & \textbf{R2} & \textbf{R3} & \textbf{NT\textsuperscript{*}} & \textbf{R1\textsuperscript{**}} & \textbf{R2} & \textbf{R3} 
      & \textbf{R1\textsuperscript{**}} & \textbf{R2} & \textbf{R3} & \textbf{R1\textsuperscript{**}} & \textbf{R2} & \textbf{R3} & \textbf{R1\textsuperscript{**}} & \textbf{R2} & \textbf{R3} \\ \hline
      
      ${SpO_2}$ 
      &  & 5.54 & 5.88 & 5.81 
      &  &  & 1.52 & 1.54
      & 69.53 & 61.15 & 84.77 
      &  & 15.81 & 22.47
      & \cellcolor[HTML]{fffabc} 69.53 & \cellcolor[HTML]{fffabc} 76.96 & \cellcolor[HTML]{fffabc} 107.24
      \\ \cline{1-1} \cline{3-5} \cline{8-12} \cline{14-18}
      
      Visualizer 
      &  & 5.55 & 5.61 & 5.46 
      &  &  & 1.41 & 1.41 
      & 59.27 & 61.93 & 90.69 
      &  & 15.57 & 23.42
      & \cellcolor[HTML]{affaaa} 59.27 & \cellcolor[HTML]{affaaa} 77.50 & \cellcolor[HTML]{affaaa} 114.11
      \\ \cline{1-1} \cline{3-5} \cline{8-12} \cline{14-18}
      
      RabbitMQ 
      &  & 5.43 & 5.62 & 5.26 
      &  &  & 1.43 & 1.47 
      & 40.67 & 34.06 & 56.86 
      &  & 8.67 & 15.89
      & \cellcolor[HTML]{fbdaff} 40.67 & \cellcolor[HTML]{fbdaff} 42.73 & \cellcolor[HTML]{fbdaff} 72.75
      \\ \cline{1-1} \cline{3-5} \cline{8-12} \cline{14-18}
      
      ECG 
      &  & 5.65 & 5.81 & 5.65 
      &  &  & 1.46 & 1.44 
      & 43.31 & 46.77 & 67.86 
      &  & 11.75 & 17.29
      & \cellcolor[HTML]{a0c5e5} 43.31 & \cellcolor[HTML]{a0c5e5} 58.52 & \cellcolor[HTML]{a0c5e5} 85.15
      \\ \cline{1-1} \cline{3-5} \cline{8-12} \cline{14-18}
      
      DB 
      & \multirow{-5}{*}{4.61} & 5.47 & 5.70 & 5.48 
      & \multirow{-5}{*}{0.80} & \multirow{-5}{*}{-} & 1.51 & 1.38 
      & 69.09 & 53.64 & 115.30 
      & \multirow{-5}{*}{-} & 14.21 & 29.04 
      & \cellcolor[HTML]{d3d3a3} 69.09 & \cellcolor[HTML]{d3d3a3} 67.85 & \cellcolor[HTML]{d3d3a3} 144.34
      \\ \hline
\end{tabular}
\end{adjustbox}

      \vspace{1px}
      NT* $\rightarrow$ no traffic, \hspace{10px} R1** $\rightarrow$ nanoservices downloaded through WiFi, 
      \hspace{10px} R2 \& R3 $\rightarrow$ nanoservices downloaded through Quectel Modem.

      \vspace{-10px}
\end{table*}

In the no-traffic situation, the RPi 5 and the Quectel modem consume 4.61 W and 0.80 W, respectively. The total energy consumption for a download event is the sum of energy consumed by RPi 5 and Quectel modem. The mean energy of the devices is calculated with equation~\ref{eqn:energy} by using the mean power and the mean download time. In the experiment, while a nanoservice is downloaded from R1, we do not consider the power consumption of the Quectel modem, as the RPi 5 uses the built-in WiFi module rather than the Quectel modem. Thus, the total energy consumption is equal to the energy consumption observed at the RPi 5 when a nanoservice is downloaded from R1.

In the experiment, to download the $SpO_2$ nanoservice, RPi 5 consumes 69.53 J of energy. While the same nanoservice is downloaded from R2 and R3, the energy consumption reaches 76.96 J and 107.24 J, respectively. As a result, R2 and R3 are causing 10.69\% and 54.24\% more energy consumption for a single download event. Similarly, R2 and R3 registries are causing more energy consumption, e.g., 30.76\% and 92.53\% for Visualizer, 5.07\% and 78.88\% for RabbitMQ, 35.12\% and 96.61\% for ECG nanoservices while these nanoservices download from the R2 and R3 registries rather than the R1 registry. The DB nanoservice, however, consumed slightly less energy while it was downloaded from the R2 compared to R1. The download from R3 increases the energy consumption by 108.92\%.

\section{Discussion and Future Work}
\label{sec:discussion-and-future-works}

\noindent Although cloud-based service registries are globally available and powerful in normal conditions, they depend on reliable network connectivity, which is a critical drawback in the deployment of critical healthcare services. This study evaluated the performance and efficiency implications of a three‑tier service deployment architecture -- addressing the above-mentioned drawback -- in a real-world use case scenario. Using a Raspberry Pi 5 (RPi 5) as a representative resource‑constrained device and an emergency ambulance scenario as the motivating use‑case, we examined how the registry placement affects the nanoservice deployment time, as well as the energy usage, network bandwidth, and computational load during the deployment.  

The performance of service deployments depends on the registry discovery from the client and the actual nanoservice download time. 
Both the RTT and number of hops in the registry discovery part gradually increase from the local registry to the remote centralized registry. 
However, the experiment revealed that the client quickly downloads a nanoservice from the local registry without requiring the external network in the three-tier registry setup. The client requires slightly more time when the same nanoservice is downloaded from an MEC-based registry rather than a remote centralized registry.

Meanwhile, we also observed that deploying the nanoservices from the local registry leads to the lowest energy usage. Deploying from the registries located further away was observed to increase energy consumption due to longer data paths and communication overhead. We conclude that registry proximity is a major factor influencing the energy consumption of a service deployment.

Moreover, our measurements related to the inflicted network load imply that the registry client requires less bandwidth when communicating with the on-premises local registry, likely due to reduced network hops and more stable local connectivity. Remote registries introduce more bandwidth overhead, partly due to external network dependencies and higher protocol negotiation costs.

Furthermore, the computational load was observed to increase with the distance between the registry and the target node, indicating that processing overhead, including network stack operations, retries, longer session handling, etc., accumulates as the registry is further away. The remote public registry performs better than the local public registry because of higher backend resources but still imposes more load on the RPi 5 than the local private registry. 

The qualitative trends suggest that registry proximity reduces device-side energy, bandwidth, and processing overhead by offering resiliency with fewer cyber attacks. Apart from this, the three-tier setup also supports scalability, where registries can offload storage and serve as alternatives. However, the three-tier registry setup has some limitations despite the various benefits. For instance, local registries require dedicated infrastructure and maintenance, which may not be feasible for all deployments. Meanwhile, MEC-based local public registries can suffer from resource bottlenecks in case they serve a high number of nodes, leading to degraded performance and higher energy and CPU overhead on end devices. 

This study focused on the service deployment. The full service lifecycle from developing nanoservices to deploying those into the devices in the hospital remains as the future work, including the criteria on selecting which service images should be made available locally. Furthermore, the devices may vary highly in terms of computational performance and capacity, as well as energy efficiency, capacity, and status, thus requiring more detailed performance and efficiency studies tailored to specific use case scenarios.

Furthermore, we see a need for developing a layered trust model to optimize the tradeoff between performance and efficiency. For example, a local private registry is generally seen as more secure than public registries since the data is processed closer to the source, reducing the exposure to external threats. On the other extreme, public registries can be considered less trustworthy due to broader exposure to potentially hostile actors, thus requiring heavier security solutions. 
The recent advancements and integration of AI/ML-based approaches to dynamically assess and maintain trust scores in real-time, considering parameters like history of security incidents and breaches, response time, and traffic anomalies, seem to be a promising area for future research.

\section{Conclusion}
\label{sec:conclusion}

\noindent This paper assessed a three‑tier registry architecture for nanoservice distribution to edge devices in an emergency ambulance context. The results show that registry proximity is the dominant factor affecting efficiency: the local registry yields the lowest energy use, bandwidth consumption, and CPU load on the Raspberry Pi 5, while MEC‑based and cloud registries introduce progressively higher overheads due to increased communication distance and processing requirements. Nevertheless, the upper‑tier registries provide scalability, resilience, and operational continuity that complement the performance advantages of the local tier. Overall, a local‑first strategy, supported by MEC and cloud fallback, offers a balanced and dependable approach for resource‑constrained, latency‑critical edge environments.

\section*{Acknowledgments}
\noindent
This research is supported by the Business Finland projects under Tomohead (grant 8095/31/2022), and Research Council of Finland-funded projects 6G Flagship (369116) and Profi6 (336449), Finnish Doctoral Program Network in Artificial Intelligence, AI-DOC (decision number VN/3137/2024-OKM-6). Ayan Mondal acknowledges the support from IIT Bhilai Innovation and Technology Foundation (grant IBITF/Note/EIR-PRAYAS/SanctionLetter/2024-25/0897).
 

\balance

\bibliographystyle{IEEEtran}
\bibliography{references}

\begin{thebibliography}{10}
\providecommand{\url}[1]{#1}
\csname url@samestyle\endcsname
\providecommand{\newblock}{\relax}
\providecommand{\bibinfo}[2]{#2}
\providecommand{\BIBentrySTDinterwordspacing}{\spaceskip=0pt\relax}
\providecommand{\BIBentryALTinterwordstretchfactor}{4}
\providecommand{\BIBentryALTinterwordspacing}{\spaceskip=\fontdimen2\font plus
\BIBentryALTinterwordstretchfactor\fontdimen3\font minus \fontdimen4\font\relax}
\providecommand{\BIBforeignlanguage}[2]{{%
\expandafter\ifx\csname l@#1\endcsname\relax
\typeout{** WARNING: IEEEtran.bst: No hyphenation pattern has been}%
\typeout{** loaded for the language `#1'. Using the pattern for}%
\typeout{** the default language instead.}%
\else
\language=\csname l@#1\endcsname
\fi
#2}}
\providecommand{\BIBdecl}{\relax}
\BIBdecl

\bibitem{9964122}
S.~M. Nagarajan, G.~G. Devarajan, A.~S. Mohammed, T.~V. Ramana, and U.~Ghosh, ``Intelligent task scheduling approach for iot integrated healthcare cyber physical systems,'' \emph{IEEE Transactions on Network Science and Engineering}, vol.~10, no.~5, pp. 2429--2438, 2023.

\bibitem{Nuguri2026188}
S.~S. Nuguri, S.~C. Bhamidipati, A.~Kambhampati, K.~Karthik, A.~Calyam, M.~K. Duvvuri, M.~L. Alarcon, and P.~Calyam, ``Security and privacy framework for cloud-based remote patient monitoring and in-place sensor-based care,'' \emph{Communications in Computer and Information Science}, vol. 2716 CCIS, p. 188 – 213, 2026.

\bibitem{harjula2019decentralized}
E.~Harjula, P.~Karhula, J.~Islam, T.~Lepp{\"a}nen, A.~Manzoor, M.~Liyanage, J.~Chauhan, T.~Kumar, I.~Ahmad, and M.~Ylianttila, ``Decentralized iot edge nanoservice architecture for future gadget-free computing,'' \emph{IEEE Access}, vol.~7, pp. 119\,856--119\,872, 2019.

\bibitem{10843208}
B.~Akdemir, H.~Faheem~Shahid, M.~A.~K. Brix, J.~Lääkkölä, J.~Islam, T.~Kumar, J.~Reponen, M.~T. Nieminen, and E.~Harjula, ``From technical prerequisites to improved care: Distributed edge ai for tomographic imaging,'' \emph{IEEE Access}, vol.~13, pp. 14\,317--14\,343, 2025.

\bibitem{11392773}
H.~F. Shahid, B.~Akdemir, J.~Islam, I.~Ahmad, I.~Ahmad, and E.~Harjula, ``Iot service orchestration in edge-cloud continuum with 6g: A review,'' \emph{IEEE Internet of Things Journal}, pp. 1--1, 2026.

\bibitem{Zerouali2018}
A.~Zerouali, ``Analyzing technical lag in docker images,'' in \emph{BENEVOL 2018}, vol. 2361, 2018, Conference paper.

\bibitem{Osorio2018}
M.~Osorio, C.~Buil-Aranda, and H.~Vargas, ``Dockerpedia: A knowledge graph of docker images,'' in \emph{ISWC (P\&D/Industry/BlueSky)}, vol. 2180, 2018, Conference paper.

\bibitem{marinescu2022cloud}
D.~C. Marinescu, \emph{Cloud computing: theory and practice}.\hskip 1em plus 0.5em minus 0.4em\relax Morgan Kaufmann, 2022.

\bibitem{9615028}
K.~Fu, W.~Zhang, Q.~Chen, D.~Zeng, and M.~Guo, ``{ Adaptive Resource Efficient Microservice Deployment in Cloud-Edge Continuum },'' \emph{IEEE Transactions on Parallel \& Distributed Systems}, vol.~33, no.~08, pp. 1825--1840, Aug. 2022.

\bibitem{9403939}
K.~Cao, S.~Hu, Y.~Shi, A.~W. Colombo, S.~Karnouskos, and X.~Li, ``A survey on edge and edge-cloud computing assisted cyber-physical systems,'' \emph{IEEE Transactions on Industrial Informatics}, vol.~17, no.~11, pp. 7806--7819, 2021.

\bibitem{9645306}
T.~Shi, H.~Ma, G.~Chen, and S.~Hartmann, ``{ Cost-Effective Web Application Replication and Deployment in Multi-Cloud Environment },'' \emph{IEEE Transactions on Parallel \& Distributed Systems}, vol.~33, no.~08, pp. 1982--1995, Aug. 2022.

\bibitem{Prabu2024277}
M.~Prabu, M.~Diviya, R.~Bhuvaneswari, D.~S. Reddy, K.~Venkatesan, and A.~K. Natarajan, \emph{The impact and integration of cloud computing for enhanced patient care and operational efficiency}.\hskip 1em plus 0.5em minus 0.4em\relax IGI Global Scientific Publishing, 2024.

\bibitem{9301260}
Z.~Ning, P.~Dong, X.~Wang, S.~Wang, X.~Hu, S.~Guo, T.~Qiu, B.~Hu, and R.~Y.~K. Kwok, ``{ Distributed and Dynamic Service Placement in Pervasive Edge Computing Networks },'' \emph{IEEE Transactions on Parallel \& Distributed Systems}, vol.~32, no.~06, pp. 1277--1292, Jun. 2021.

\bibitem{10614854}
Z.~Xiang, Y.~Zheng, D.~Wang, J.~Taheri, Z.~Zheng, and M.~Guo, ``{ Cost-Effective and Robust Service Provisioning in Multi-Access Edge Computing },'' \emph{IEEE Transactions on Parallel \& Distributed Systems}, vol.~35, no.~10, pp. 1765--1779, Oct. 2024.

\bibitem{10129089}
G.~Cui, Q.~He, X.~Xia, F.~Chen, and Y.~Yang, ``{ EESaver: Saving Energy Dynamically for Green Multi-Access Edge Computing },'' \emph{IEEE Transactions on Parallel \& Distributed Systems}, vol.~34, no.~07, pp. 2155--2166, Jul. 2023.

\bibitem{Suriyan2025323}
K.~Suriyan, S.~Ganesh, P.~Palaniyammal, G.~Priya, and V.~Singh, \emph{Recent Trends in Edge Computing: Challenges and Opportunities}.\hskip 1em plus 0.5em minus 0.4em\relax IGI Global Scientific Publishing, 2025.

\bibitem{Hussain2024388}
N.~Hussain, V.~Dankan~Gowda, C.~Shyamsunder, V.~Srinivas, R.~Rani, and M.~Balaji, ``Optimizing iot device networks with edge computing to address latency and bandwidth constraints,'' in \emph{2024 5th international conference on electronics and sustainable communication systems (ICESC)}, 2024, Conference paper, p. 388 – 395.

\bibitem{Ramanathan20241}
S.~Ramanathan, A.~Pineda-Briseno, T.~K. Mohd, and M.~Ramasundaram, \emph{Edge Computing in Healthcare: Concepts, Tools, Techniques, and Use Cases}.\hskip 1em plus 0.5em minus 0.4em\relax CRC Press, 2024.

\bibitem{Karami20256183}
A.~Karami and M.~Karami, ``Edge computing in big data: challenges and benefits,'' \emph{International Journal of Data Science and Analytics}, vol.~20, no.~7, p. 6183 – 6226, 2025.

\bibitem{8931321}
J.~Islam, E.~Harjula, T.~Kumar, P.~Karhula, and M.~Ylianttila, ``Docker enabled virtualized nanoservices for local iot edge networks,'' in \emph{2019 IEEE Conference on Standards for Communications and Networking (CSCN)}, 2019, pp. 1--7.

\bibitem{shahid2024resource}
H.~F. Shahid and E.~Harjula, ``Resource slicing through intelligent orchestration of energy-aware iot services in edge-cloud continuum,'' in \emph{Proceedings of the 14th International Conference on the Internet of Things}, 2024, pp. 244--245.

\bibitem{11220483}
H.~F. Shahid, J.~Islam, I.~Ahmad, and E.~Harjula, ``Optimizing resource-aware service orchestration in edge-cloud continuum,'' in \emph{2025 IEEE Intelligent Mobile Computing (MobileCloud)}, 2025, pp. 44--50.

\bibitem{wiig2021resilient}
S.~Wiig and J.~K. O’Hara, ``Resilient and responsive healthcare services and systems: challenges and opportunities in a changing world,'' \emph{BMC Health Services Research}, vol.~21, pp. 1--5, 2021.

\bibitem{liu2021performance}
Q.~Liu, K.~G. Mkongwa, and C.~Zhang, ``Performance issues in wireless body area networks for the healthcare application: a survey and future prospects,'' \emph{SN Applied Sciences}, vol.~3, pp. 1--19, 2021.

\bibitem{distributedorchestration}
A.~Pires, J.~Sim{\~a}o, and L.~Veiga, ``Distributed and decentralized orchestration of containers on edge clouds,'' \emph{Journal of Grid Computing}, vol.~19, pp. 1--20, 2021.

\bibitem{9903191}
U.~C. Özyar and A.~Yurdakul, ``A decentralized framework with dynamic and event-driven container orchestration at the edge,'' in \emph{2022 IEEE International Conferences on Internet of Things (iThings) and IEEE Green Computing \& Communications (GreenCom) and IEEE Cyber, Physical \& Social Computing (CPSCom) and IEEE Smart Data (SmartData) and IEEE Congress on Cybermatics (Cybermatics)}, 2022, pp. 33--40.

\end{thebibliography}

\vfill

\end{document}